\documentclass[12pt]{article}
\usepackage{amsmath,amssymb}
\usepackage[pdftex]{graphicx}
\usepackage{hyperref}

\begin{document}

\title{\bf Incoherent light as a control resource:\\ a route to complete
controllability of quantum systems}

\author{Alexander Pechen\footnote{Department of Chemical Physics,
Weizmann Institute of Science, Rehovot 76100, Israel; Steklov Mathematical
Institute of Russian Academy of Sciences, Gubkina 8, Moscow 119991,
Russia; Email:~apechen@gmail.com;
\href{http://www.mathnet.ru/eng/person17991}{
http://www.mathnet.ru/eng/person17991}}}
\date{}
\maketitle

\vspace{-.8cm}
\begin{abstract}
We discuss the use of incoherent light as a resource to control the atomic
dynamics and review the proposed in Phys. Rev. A 84, 042106 (2011) method for
a controlled transfer between any pure and mixed states of quantum systems using
a combination of incoherent and coherent light. Formally, the method
provides a constructive proof for an approximate open-loop Markovian
state-transfer
controllability of quantum system in the space of all density
matrices---the strongest possible degree of quantum state control.
\end{abstract}

\section{Introduction}
Optical control of quantum dynamics plays a key role in quantum
information and quantum technology~\cite{Sergienko2006}. Commonly manipulation
by quantum systems is
realized using coherent shaped
laser pulses~\cite{Tannor1985,Brumer2003,Letokhov2007,Zhdanov2008,Steinitz2012}.
In recent years
various methods exploiting incoherent environments as control resources for
quantum information and quantum technology were
suggested~\cite{Pechen2006,Pechen2008,Cirac2009,Weimer2010,Blatt2010,
Rossini2007}. In this work we discuss the
potential for using incoherent light as a control resource for manipulating
quantum systems as proposed in~\cite{Pechen2006,Pechen2008} and review the
method for a controlled
deterministic open-loop transfer between arbitrary pure and mixed states
using a special
combination of incoherent and coherent photons~\cite{Pechen2011}---a result
having a fundamental interest as proving the possibility of achieving
the maximal degree of
quantum state control and practical interest for possible applications in
quantum computing
with mixed states~\cite{Aharonov1997,Tarasov2002}, where engineering of
arbitrary density matrices is a necessary component.

Section~2 of this work discusses the notions of coherent and incoherent
controls; the former is commonly realized by shaped laser pulses while the
latter by
tailored state of the environment (for example, by spectral density of
incoherent light). Section~3 outlines the proposed physical scheme for
steering an arbitrary initial atomic pure or mixed state into vicinity of any
final pure or mixed state. This scheme includes two stages, incoherent
(described in Section~3.1) and coherent (described in Section~3.2) and proposes
a laboratory realization of controllability of open quantum systems in the space
of all density matrices. Both stages are equally important for the efficient
work of this scheme. Section~3.3 considers as an example the application of this
scheme for preparation of a special mixture of $4^1 S$ and $4^1 P$ states of
Ca atoms. Section~4 explains why our scheme practically proves complete
controllability of Markovian quantum systems in the space of all density
matrices.

\section{Coherent and incoherent controls}
Consider an open, that is, interacting with an environment, $n$-level quantum
system evolving
under the action of a shaped laser pulse $u(t)$. Quantum mechanical state of the
system at time $t$ is described by a density matrix $\rho_t$ which is a positive
matrix with unit trace. Let ${\cal D}_n=\{\rho\in\mathbb C^{n\times n}\,|\,
\rho\ge 0, {\rm
Tr}\rho=1\}$, where ${\rm Tr}$ denotes trace, be the set of all density
matrices for an $n$-level quantum system.

On a sufficiently short time scale the decoherence effects can be neglected and
the laser pulse induces the unitary system dynamics governed by the equation
\begin{equation}\label{eq1}
\dot\rho_t=-i\left[H_0+Vu(t),\rho_t\right]
\end{equation}
Here $H_0=\sum\limits_{i=1}^n\varepsilon_i |i\rangle\langle i|$ is the free
Hamiltonian of the system
($\varepsilon_1\le\varepsilon_2\le \dots\le\varepsilon_n$) and $V=-\mu$,
where $\mu$
is the dipole moment.

On the long time scale the influence of the environment becomes essential.
Commonly, the environment is considered as fixed and having
deleterious effects on the system, and moreover its action on the system is
assumed to be out
of our control. However, this assumption is too restrictive since we can
manipulate
the state of the environment by adjusting its temperature, pressure or, more
generally, its distribution function $n_\omega$. Here we consider a simple case
when the environment is characterized by the distribution of its particles in
energy $\omega$. If the environment is Markovian, then the reduced dynamics of
the system density matrix will be governed by the master equation
\begin{equation}\label{eq2}
\dot\rho_t=-i\left[H_0+H_{n_\omega},\rho_t\right]+{\cal L}_{n_\omega}(\rho_t)
\end{equation}
where the Lindblad superoperator ${\cal L}_{n_\omega}$ is determined by the type
of the environment, its state ${n_\omega}$, and the microscopic details of
its interaction with the system. One particularly convenient environment is
spectrally filtered incoherent light~\cite{Pechen2006}. The Lindblad
superoperator induced by incoherent light with spectral density ${n_\omega}$
(the distribution of photons in frequency $\omega$) has the form
\[
{\cal L}_{n_\omega}(\rho)=\sum\limits_{i<j} A_{ij}
\left[(n_{\omega_{ij}}+1)L_{Q_{ij}}(\rho)+n_{\omega_{ij}}L_{Q_{ji}}
(\rho)\right ]
\]
where $A_{ij}\ge 0$ are the Einstein coefficients for spontaneous emission,
$\omega_{ij}=\varepsilon_i-\varepsilon_j$ are the system transition frequencies,
$Q_{ij}=|i\rangle\langle j|$ is the transition operator for the
$|j\rangle\to|i\rangle$ transition, and $L_Q(\rho)=2Q\rho Q^\dagger-Q^\dagger
Q\rho-\rho Q^\dagger Q$. The intensities $u(t)$ and $n_\omega$ of the
coherent and incoherent light are the coherent and incoherent controls,
respectively.

\section{State transfer between arbitrary pure and mixed quantum states}
Here we discuss the scheme were incoherent light with some spectral density
$n_\omega$ is used as a control resource together with a suitable coherent laser
field
$u(t)$ to approximately implement state transfer between arbitrary
pure and mixed states of
finite-level quantum systems~\cite{Pechen2011}. The scheme is described below
for implementing state transfer to non-degenerate density matrices,
i.e., density matrices
with all distinct eigenvalues. Such density matrices are dense in the set ${\cal
D}_n$
of all density matrices and therefore any pure or mixed quantum state can
be approximated with arbitrary precision by non-degenerate states. Thus state
transfer to such states is sufficient for any practical purpose.

Let $\rho_{\rm f}=\sum p_i |\phi_i\rangle\langle\phi_i|$ be any
desired non-degenerate (i.e., $p_i\ne p_j$ for $i\ne j$) density matrix. Here
$p_i$ are its eigenvalues and $|\phi_i\rangle$ the corresponding
eigenvectors. We assume without loss of the generality that
$p_1>p_2>\dots>p_n$. The scheme for steering arbitrary initial density matrix
$\rho_{\rm i}$ into $\rho_{\rm f}$ involves the two
stages:

\begin{itemize}
\item First stage (incoherent): steering $\rho_{\rm i}$ to the state
$\tilde\rho_{\rm
f}=\sum p_i|i\rangle\langle i|$ using spectrally filtered incoherent light. The
state $\tilde\rho_{\rm f}$ has the same eigenvalues as the target state
$\rho_{\rm f}$ but is diagonal in the basis of eigenvectors of $H_0$. Note also
that the populations of the energy levels are arranged in the decreasing order
since $p_1>p_2>\dots>p_n$.
\item Second stage (coherent): unitary rotation of the basis $\{|i\rangle\}$
to match the basis $\{|\phi_i\rangle\}$ of $\rho_{\rm f}$ using coherent laser
field. This stage is standard in coherent control.
\end{itemize}

\subsection{Incoherent stage}
For this stage we assume that all system transition frequencies
$\omega_{ij}$ are different and all $A_{ij}\ne 0$. These assumptions represent
the case of general position in the sense that randomly chosen $H_0$ and
$A_{ij}$ with probability one will satisfy these assumptions. (However, as often
might happen with the case of general position there may be important physical
examples which do not fit these assumptions.) Under these assumptions any
pair of the system states can be independently addressed by incoherent light.
During this stage we switch off coherent control so that $u(t)=0$ and apply
incoherent control with the optimal spectral density $n_\omega$ which has the
constant value $n_{\omega_{ij}}=p_j/(p_i-p_j)$ in each frequency range of
significant absorption and emission for every system transition frequency
$\omega_{ij}$.
This incoherent control will exponentially fast drive \textit{any} initial
system density matrix $\rho_{\rm i}$ to $\tilde\rho_{\rm f}$:
\[
 \rho_{\rm i}\to\rho_t\to\tilde\rho_{\rm f} \qquad (t\to\infty)
\]
While the complete steering is generally achieved in infinite time, because of
the
exponential convergence the time period of several magnitudes of the
characteristic relaxation time $\tau_{\rm rel}$ is always practically
sufficient to be in a desired close proximity of $\tilde\rho_{\rm f}$. We apply
the control $n_\omega$ during such a period and then proceed to the coherent
stage.

\subsection{Coherent stage}
This stage is a standard unitary rotation. The system is assumed to be unitary
controllable on a short w.r.t. $\tau_{\rm rel}$ time scale (i.e., such that any
unitary evolution operator $U$ can be produced by some fast control). The
conditions for unitary controllability of quantum systems are
well-known~\cite{Ramakrishna1995}. Incoherent control is switched off during
this stage (i.e., $n_\omega\equiv 0$) and a fast coherent control with optimal
laser pulse $u(t)$ is applied to produce a unitary operator $U_*$ transforming
the basis $\{|i \rangle\}$ into $|\phi_i\rangle$, i.e. such that
$U_*|i\rangle=|\phi_i\rangle$. This stage is realized on a short w.r.t.
$\tau_{\rm rel}$ time scale when the decoherence effects are negligible. The
system state will evolve as:
\[
 \tilde\rho_{\rm f}\to U^{\vphantom{\dagger}}_*\tilde\rho_{\rm f}
U^\dagger_*=\rho_{\rm f}
\]
As the result of this evolution the system state is steered in the desired
state $\rho_{\rm f}$. This dynamics is approximate since even for $n_\omega=0$
the spontaneous emission will induce some non-unitary effects. However, these
effects are negligible on a sufficiently short w.r.t. $\tau_{\rm rel}$ time
scale
where unitary dynamics serves as a good approximation. This transformation
finishes steering $\rho_{\rm i}$ to $\rho_{\rm f}$.

\subsection{Example: Calcium atom}
As a testing example we numerically studied in~\cite{Pechen2011}
application of control by incoherent light for engineering a mixed state of two
calcium levels $|0\rangle=\rm 4^1 S$ and $|1\rangle=\rm 4^1 P$. The
corresponding transition frequency is $\omega_{21}=4.5\times 10^{15}$~rad/s, the
radiative lifetime $t_{21}=4.5$~ns, the Einstein coefficient
$A_{21}=1/t_{21}\approx 2.2\times 10^8$~s$^{-1}$, and the dipole moment
$\mu_{12}=2.4\times 10^{-29}$ C$\cdot$m. The method works for
any initial and target states; we choose $\rho_{\rm i}=|0\rangle\langle 0|$ and
$\rho_{\rm f}=\frac{1}{4}|0\rangle\langle 0|+\frac{3}{4}|1\rangle \langle 1|$
with greater population in the upper state.

During the first (incoherent) stage we prepare the mixture $\tilde\rho_{\rm
f}=\frac{3}{4}|0\rangle\langle 0|+\frac{1}{4}|1\rangle\langle 1|$ by applying
incoherent light with spectral density $n_{\omega_{21}}=1/2$ during the time
interval $T=50$~ns. The second (coherent) stage is realized by applying a
resonant $\pi$-pulse $E(t)=E\cos(\omega_{12}t)$ with amplitude $E=10^7$~V/m
acting during the time interval $T_{\rm f}-T=1310$~fs (other coherent control
methods producing the same unitary transformation can be used as well); this
field has the Rabi frequency $\Omega_{\rm R}\approx 1/2320$~fs$^{-1}$ and acts
as a $\pi$-pulse transforming the state $\tilde\rho_{\rm f}$ into $\rho_{\rm
f}$. The results of the numerical simulation are shown in Fig.~1.

\begin{figure}[htbp]
  \centering
  \includegraphics[scale=.69]{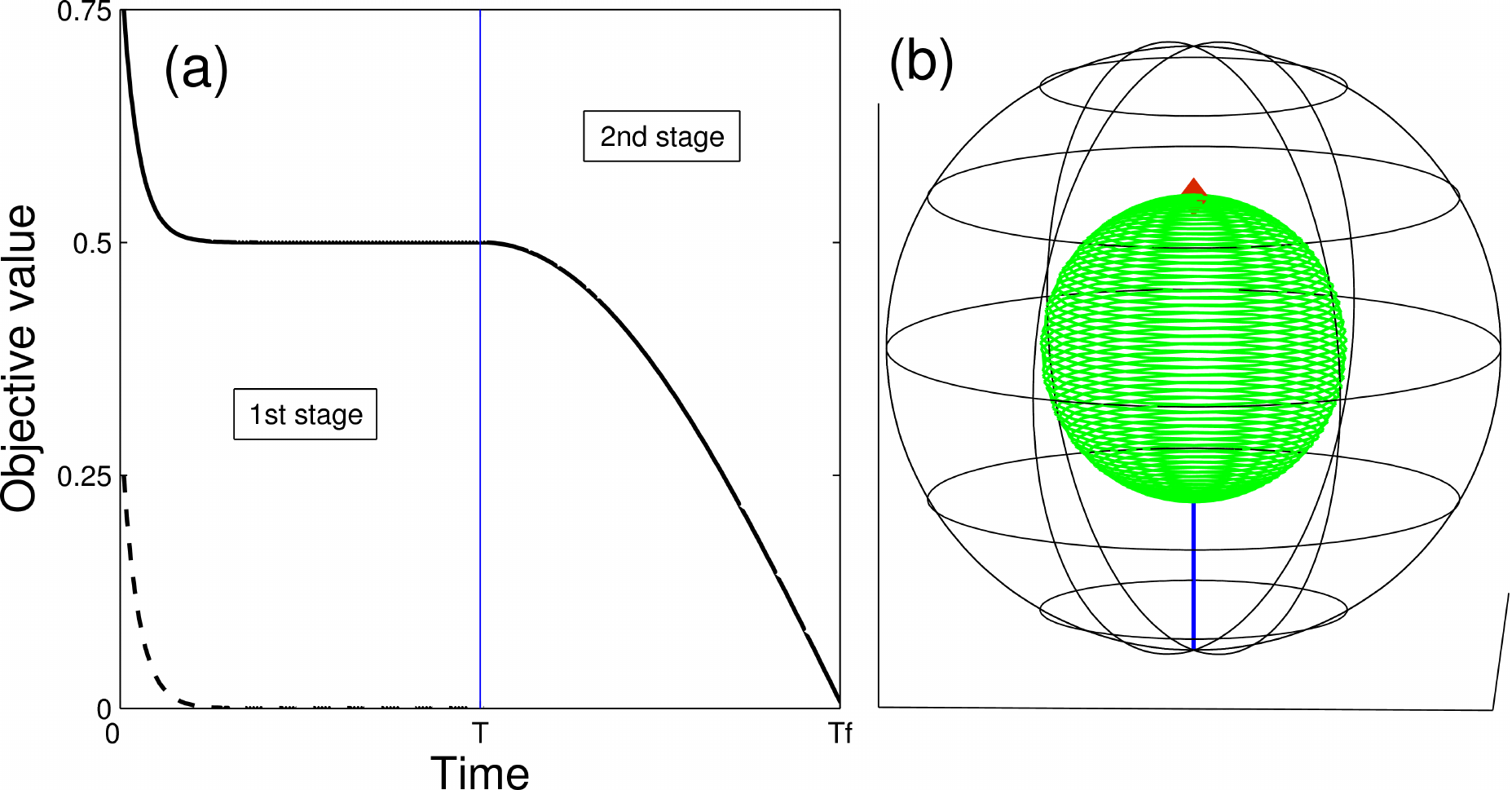} % 8.3cm
\caption{(From Ref.~\cite{Pechen2011}. Copyright (2011) by the American Physical
Society.) The
results of the numerical simulations for engineering a target mixed state
$\rho_{\rm f}=\frac{1}{4}|0\rangle\langle 0|+\frac{3}{4}|1\rangle \langle 1|$ of
two Ca energy levels. (a) The behavior of $\|\rho_t-\rho_{\rm f}\|$ (solid line)
and $\|\rho_t-\tilde\rho_{\rm f}\|$ (dashed line) during the two control
stages. The time scale is not uniform: $T=50$ ns and $T_{\rm f}-T=1310$ fs.
(b) The corresponding evolution of the Bloch vector showing that the evolved
state approaches the target state (red marker at the top).}
\end{figure}

\section{Complete controllability of quantum systems}
A fundamental property of any control system is the degree of its
controllability. The state of the system is a collection of the system's
variables which completely describes the system at any given time. Complete
state controllability (or simply controllability)
describes the ability to steer with the available controls any initial
state of the system into any final state. For quantum mechanical systems the
state space can be either the set of all pure states, or any complete set of
density matrices with the same eigenvalues (i.e., any complete set of
kinematically equivalent density matrices), or the set of all density matrices
${\cal D}_n$. The latter set is the biggest and contains as subsets the two
former sets. Therefore state controllability in ${\cal D}_n$ is the strongest
amongst all notions of quantum state control; controllability of a given quantum
system in ${\cal D}_n$ implies its controllability in the space of pure states
and in the space of kinematically equivalent density matrices.

The method of~\cite{Pechen2011} which is described in section~3 of
this paper allows to steer an arbitrary initial state into a vicinity of an
arbitrary final state. While the definition of controllability requires exact
steering, for practical purposes it is always sufficient to steer the system
into some vicinity of the final state. Thus our method practically
proves the strongest degree of state controllability of quantum
systems---controllability in the space of all density matrices ${\cal D}_n$. A
powerful method for proving controllability of bilinear systems is the Lie
algebraic method. The proof in~\cite{Pechen2011} is different: we prove
controllability by explicitly constructing optimal controls which steer any
initial state into any given final state. The success in finding these controls
is due to a special structure of the Markovian master equation describing
interaction of a quantum system with incoherent light. An important feature of
our method is that we use a physical environment comprised of incoherent photons
and its spectral density is accessible to manipulate. Another feature is that
the proposed method does not use simultaneous control by coherent and incoherent
fields. This property is important because deriving a correct master equation
with simultaneous time dependent Hamiltonian and a dissipative term is a
non-trivial problem. Simple addition of ${\cal L}_{n_\omega}$ to the r.h.s. of
eq.~(\ref{eq1}) may lead to an incorrect equation because time dependent
Hamiltonian can also modify the non-Hamiltonian part ${\cal L}_{n_\omega}$. In
our method both controls are separated in time that allows to avoid this
problem and assures that we use a correct description of the dynamics. Recently
several works have appeared which also analyze and prove controllability of
quantum systems controlled by a coherent control with a Markovian
environment. The preprint~\cite{Rabitz2012} uses a
Lie algebraic technique and the preprint~\cite{Thomas2012} uses coherent control
and a switchable noise to transfer between arbitrary states of $n$-level quantum
systems.

\section*{Acknowledgments}
This work was supported by a Marie Curie International Incoming Fellowship
within the 7th European Community Framework Programme.

\end{document}